# PSIM: A TOOL FOR ANALYSIS OF DEVICE PAIRING METHODS

Yasir Arfat Malkani and Lachhman Das Dhomeja

Department of Informatics, University of Sussex, Brighton, UK
{y.a.malkani, l.d.dhomeja}@sussex.ac.uk

*ABSTRACT*

*Wireless networks are a common place nowadays and almost all of the modern devices support wireless communication in some form. These networks differ from more traditional computing systems due to the ad-hoc and spontaneous nature of interactions among devices. These systems are prone to security risks, such as eavesdropping and require different techniques as compared to traditional security mechanisms. Recently, secure device pairing in wireless environments has got substantial attention from many researchers. As a result, a significant set of techniques and protocols have been proposed to deal with this issue. Some of these techniques consider devices equipped with infrared, laser, ultrasound transceivers or 802.11 network interface cards; while others require embedded accelerometers, cameras and/or LEDs, displays, microphones and/or speakers. However, many of the proposed techniques or protocols have not been implemented at all; while others are implemented and evaluated in a stand-alone manner without being compared with other related work [1]. We believe that it is because of the lack of specialized tools that provide a common platform to test the pairing methods. As a consequence, we designed such a tool. In this paper, we are presenting design and development of the Pairing Simulator (PSim) that can be used to perform the analysis of device pairing methods.*

## 1. INTRODUCTION

More and more computing devices are coming into existence everyday, which may vary in size, capabilities, mode of interaction and so on. As a result, we are moving towards a world in which computing is omnipresent. Most of the modern devices support multiple communication channels and almost all of them use wireless technology in some form, such as Bluetooth, Infrared, Wibree, Zigbee, Ultrasound or 802.11 Having wireless technology in these devices does not guarantee that all of these devices can also take advantage of internet technology. However, those wireless enabled devices that can not be connected to internet, can still take advantage of other co-located devices in the vicinity by forming short-term or long-term associations on ad hoc basis. For example pairing a Bluetooth enabled headset with a mobile phone or MP3 player (short-term) and pairing of a PDA with home devices in order to control them wirelessly (long-term).

Since wireless communication is susceptible to eavesdropping, thus one can easily launch well known man-in-the-middle (MITM) attack, Denial-of-Service (DoS) attack or can perform bidding-down attack to fail the secure pairing process. The solution to this problem is establishing a secure channel between the pairing devices by some kind of mechanisms, such as authentication and encryption. Establishing a secure channel is trivial, when there existed an off-line or on-line global infrastructure, such as PKI. However, such a global infrastructure is hard to implement in mobile ad hoc and ubiquitous computing environments that makes it a challenging real-world problem. Due to the ad hoc and dynamic nature of these environments, devices do not know each other a priori, so the idea of pre-shared secret key is failed. Further, traditional key exchange or key agreement approaches, such as Diffie-Hellman [2]– in their actual form – are not applicable in wireless environments due to their vulnerability to MiTM attack. Moreover, devices' heterogeneity in terms of their communication channels, user interfaces, power requirements and sensing technology, make it hard to give a single solution for secure pairing of devices. As a result, wide community of researchers has proposed a large set of protocols and techniques to deal with this issue. However, these protocols vary in the strength of their security, their susceptibility to environmental conditions and in the required physical capabilities of the devices. Currently, there may existed many options for an ordinary user to establish a secure channel between the devices from entering pins and passwords to verifying hashes of public keys and pressing buttons simultaneously on the two devices. In this paper, we are presenting the design, development and evaluation of a simulation tool for pairing methods followed by a brief discussion on existing pairing methods.

Remaining part of this paper is organized as follows: section 2 is background that describes various existing device pairing methods, section 3 presents the design and development of the simulator, section 4 discusses a case study that is carried out to evaluate the simulator, and finally section 5 concludes the paper.

## 2. BACKGROUND

The problem of secure device association (pairing) continues to be a very active area of research in wireless environments. The issue got significant attention from many researchers, after Stajano et al. [3-4] highlighted the challenges inherent in secure device association. They proposed a mother-duckling (master-slave) model, which maps the relationship between devices. The pairing process is done by agreeing on a secret key over the physical connection (such as using a cable). Though the secret key is transferred in plain-text and cryptographic methods are not used, it is susceptible to dictionary attacks. In reality, it is also difficult to have common physical interfaces in both of the devices, and carrying cables all of the time might not be possible for owners of the devices. Balfanz et al. [5] extended Stanjano and Anderson's work and proposed a two-phase authentication method for pairing of co-located devices using infrared as a location limited side channel. In their proposed solution, pre-authentication information is exchanged over the infrared channel and then the user switches to the common wireless channel. Slightly different variations of Balfanz et al [5] approach are proposed in [6-9], which also use location limited side channel to transfer the pre-authentication data. The common problem with these approaches is twofold: first, they need some kind of interface (e.g. IrDA, laser, ultrasound, etc) for pre-authentication phase and are vulnerable to passive eavesdropping attack in the location limited side channels, e.g. two remotes and one projector. Some other pairing schemes including Bluetooth require the human operator to put the communicating partners into discovery mode. After discovery and selection of a device, the channel is secured by entering the same PIN or password into both devices. Although it is a general approach, it gives rise to a number of usability and security issues [10, 11]. For example, a short password or PIN number makes it vulnerable to dictionary or exhaustive search attacks. Further, in Bluetooth pairing an adversary can eavesdrop to break the security from a long distance using powerful antennas.

Based on the pairing protocol of Balfanz et al. [5], some other schemes are proposed through the use of audio and visual out-of-band channels. One such system is Seeing-is-Believing (SiB) [12]. SiB uses two dimensional bar codes for exchanging pre-authentication data between the devices. In the proposed approach, device A encodes cryptographic material into a two-dimensional barcode and displays it on the screen, then device B reads it through a camera to setup an authenticated channel. To reduce the camera requirement in one of the pairing device in SiB, Saxena et al. [13] extended the work of McCune et al. [12] and proposed an improvement to it through the use of simple light source, such as LEDs, and short authenticated integrity checksums. In the proposed scheme, device A needs to be equipped with a camera and device B with a single LED. When the LED on device B blinks, device A takes a video clip. Then, video clip is parsed to extract an authentication string. Loud and Clear (L&C) [14] and Human-Assisted Pure Audio Device Pairing (HAPADEP) [15] use audio as an out-of-band channel to securely pair the devices. The main idea of L&C [14] scheme is to encode the hash of first device's public key into a MadLib sentence (i.e. grammatically correct but nonsensical sentence) and transmit it over a device-to-human channel using a speaker or a display. Then, second device also encodes the hash of the received public key from first device into the MadLib sentence and transmit it over a device-to-human channel using a speaker or a display. Then user is responsible to compare the two sentences and to accept or reject the pairing. There are two variants of this approach: speaker-to-speaker and display-to-speaker. In first method user is required to compare and verify the two sentences vocalized by the pairing candidate devices. In the second method, user is required to compare the displayed MadLib sentence on one device with the vocalized MadLib sentence from the other device. Finally, user is responsible for accepting or rejecting the pairing based on the results of comparison. In HAPADEP [15], Soriente et al. consider the problem of pairing two devices that have no common wireless communication channel, such as Bluetooth or WiFi, at the time of pairing. The proposed scheme uses pure audio to exchange both public keys and hashes of public keys. The pairing schemes that use audio and/or visual out of band channels [12-14] are also inapplicable in some of the scenarios. For example, SiB [12] requires that devices must be equipped with camera; while in L&C [14] a speaker and/or display is required, and HAPADEP [15] is applicable in those scenarios where both devices have a microphone and a speaker. Camera equipped devices are usually prohibited in high security areas; while the latter is not suitable for hearing-impaired users. Further, bar code scanning requires sufficient proximity and light in SiB; while L&C and HAPADEP places some burden on the user for comparison of audible sequences. An adversary can easily subvert bar code stickers on devices in SiB; while ambient noise makes authentication either weak or difficult in L&C as well as in HAPADEP. While [13] is a variation of SiB, so this scheme has few of the same limitations as SiB, such as requiring close proximity and a camera in at least one of the device.

Unlike previously described approaches, the idea of shaking devices together to pair them has become more common. Smart-its-Friends [16] is the first effort that proposed pairing of two devices using a common movement pattern and used accelerometers as an out-of-band channel. In this approach, two devices are held and shaken together simultaneously. Then, common readings from the embedded accelerometers in the devices are exploited to establish the communication channel between the two devices. However, security has not been the major concern of Smart-its-Friends. The follow-on method to Smart-its-Friends is Shake Well Before Use [17]. Mayrhofer and Gellersen extended the Holmquist et al. [16] approach and proposed two protocols to securely pair the devices. Both of the proposed protocols exploit the cryptographic primitives with accelerometer data analysis for secure device-to-device authentication. First protocol uses public key cryptography and is more secure as compared to the second protocol, which is more efficient and computes secret key directly from accelerometer's data. Another approach that requires shaking or moving patterns is Shake Them Up [18]. Authors suggest a manual technique for pairing two resource-constrained devices that involves shaking and twirling them in very close proximity to each other. Unlike Smart-its-Friends and Shake Well Before Use, this approach exploits the source indinguishability property of radio signals and does not require embedded accelerometers. While being shaken, two devices exchange radio packets and agree on a key one bit at a time, relying on the adversary's inability to determine the source of radio packet (sending device). Secure pairing of devices by shaking devices together is an interesting approach. However, these schemes require 2D accelerometers in both of the devices. Further, shaking devices together is always not possible, since there is large variety of devices, such as printers, projectors and laptops that can not be hold and shaken together simultaneously. While Shake Them Up is susceptible to attack by an eavesdropper that exploits the differences in the baseband frequencies of the two radio sources. Recently, Varshavsky et al. [19] proposed – Amigo – a proximity-based technique for secure pairing of co-located devices. Authors extended Diffie-Hellman key exchange protocol with the addition of key verification stage. The proposed approach exploits the common radio signals from locally available wireless access points to establish the secure channel between the devices. Since, AMIGO uses the common radio signals from the located access points; it is not applicable in the scenarios, where the radio data is not available to process or where the wireless network is easy to eavesdrop on while remaining hidden. It is also a fact that in many under developing countries 802.11-based wireless technology is less popular as compared to Bluetooth technology that is more popular and common due to the mobile phones.

Some other efforts towards secure device pairing include Button-Enabled Device Association (BEDA) [20], LoKey [21], Are You With Me? [22] and Malkani et al.'s work [23, 24]. BEDA is proposed by Soriente et al., and it has four variants. The main idea is to transfer the short secret key from one device to the other using 'button-presses' and then use that key to authenticate the public keys of the devices. Short secret key is agreed upon between the two devices via one of the four variants of BEDA. These are called button-to-button (B-to-B), display-to-button (D-to-B), short vibration-to-button (SV-to-B) and long vibration-to-button (LV-to-B). The first and basic variant (i.e. B-to-B) involves the user simultaneously pressing buttons on both of the devices within certain random time-intervals and each of these intervals are used to derive 3-bits of the short secret key. LoKey uses SMS messages to authenticate key exchanged over the internet. However, this approach incurs substantial monetary cost and delay. While, Are You With Me? again requires accelerometers and is not applicable in the scenarios as Shake Well Before Use or Smart-its-Friends. Malkani et al. [23, 24] have proposed a generic framework for secure device association. In the proposed system devices first register their capabilities with the directory service. Then, whenever two devices need to create an association, the client (device A) queries the directory service to discover and acquire the required information to initiate a secure pairing with the target device (device B). Based on the information from directory service, both the client (device A) and resource (device B) mutually execute a common pairing protocol. This protocol involves the generation of a key from interaction with the environment. The selected interactions generate an appropriate key for the nature of the intended association, and a successful pairing arises when matching keys are generated on both of the devices.

In summary, there is an immense literature on secure device association. However, some of the proposed techniques or protocols are not implemented at all; while others are implemented and evaluated in a stand-alone manner without being compared with other related work [1, 25]. Examples of these include Resurrecting Duckling Security Model [4], Talking to Strangers [5], AMIGO [19], Shake Well Before Use [17], some of the Saxena et al.'s proposed methods [26] and four variants of BEDA [20] approach. Since the motivation for this work is [25] that presents state-of-the-art in the area of secure device pairing along with some future research directions, therefore, one can refer it for further details.

## 3. DESIGN AND DEVELOPMENT

An extensive set of comparative usability tests can play a vital role in the process of standardizing pairing methods. However, conducting such a detailed case study is not an easy task due to several reasons. For example, it is very tedious and laborious job to implement all of the existing (more than two dozens) pairing methods using a common platform. Even it becomes more difficult when implementing network functionality since these schemes use numerous wireless channels, such as Bluetooth, WiFi, Ultrasound, Infrared, etc. We believe that

International Journal of Network Security & Its Applications (IJNSA), Vol.1, No.3, October 2009

our designed simulator reduces the development and implementation efforts for pairing protocols and makes it easy to conduct several usability tests to evaluate them.

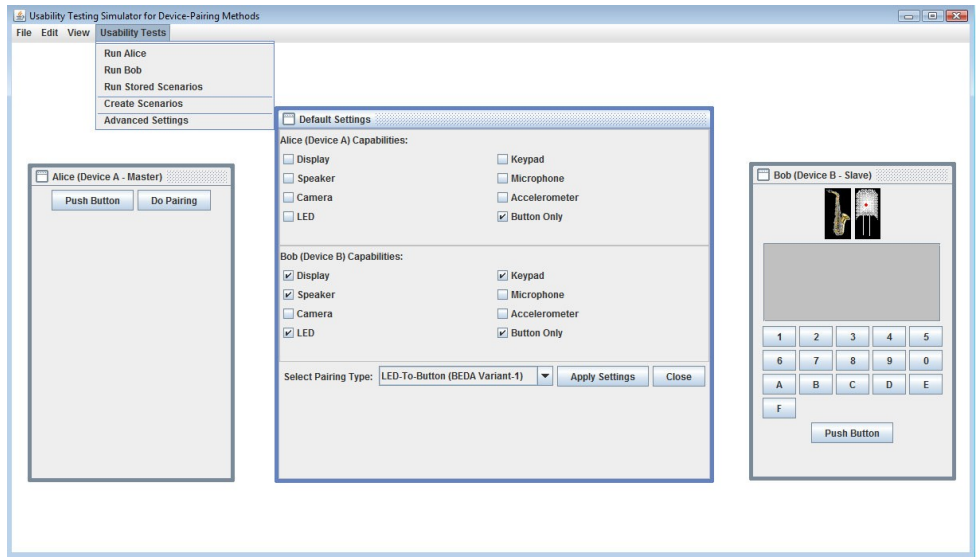

Figure 1: Screen shot of the simulator showing simulated devices (Alice and Bob)

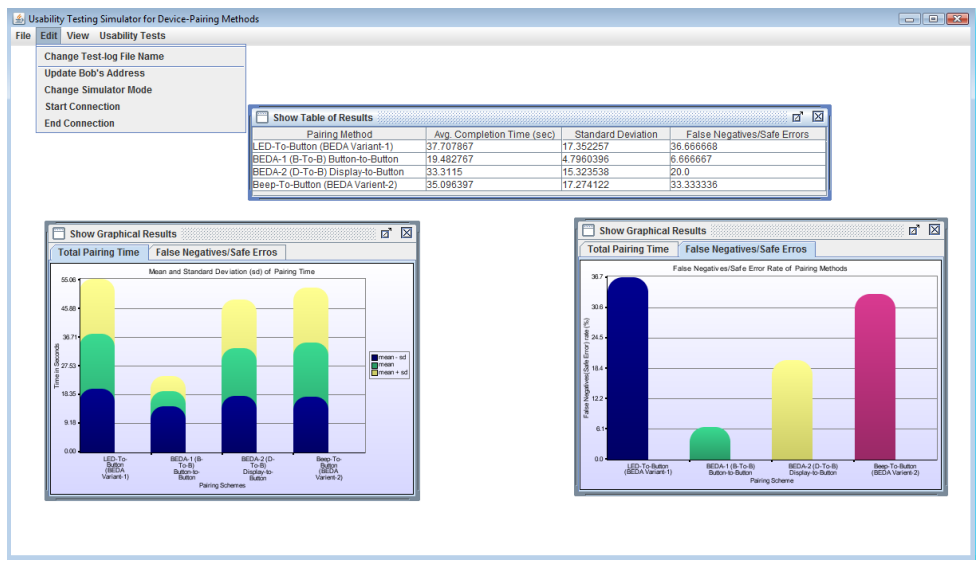

Figure 2: Screen shot of the simulator showing some of the test results

Figure 1 and figure 2 shows the screen shots of the simulator. Our simulator supports automated test sessions, automatic test data collection, logging errors, and also facilitates to simulate several attack scenarios, such as MiTM. Test organizer/developer does not need to develop new device interfaces for pairing methods each time, since the simulator supports simulation of devices having combinations of eight installed capabilities. Therefore, existing and new pairing methods can easily be implemented and tested with minimum development efforts. Figure 1 illustrates a typical test session, while figure 2 illustrates how does test organizer can see the results at the end of the test without taking any extra efforts. The designed simulator is capable of showing final results in tabular as well as in graphical/charts form. We have used java to write the code of simulator. Eclipse is used as an IDE. Some of the third party libraries used in the development of this simulator include Bouncy Castle Cryptographic library [27] and Chart2D library [28]. Bouncy Castle is a lightweight collection of APIs used in cryptography. These APIs can work with J2ME, J2SE and there are also APIs for C# programming language. While Chart2D is a Java class library for drawing 2D charts or graphs. Some of the features of the designed simulator are summarized below:

1. Simulator can be used either locally running on one computer, or remotely running on two different computers (i.e. one for Alice and other for Bob)

2. It facilitates logging of test date and time, pairing method's parameters (such as pairing scheme name, total pairing time, false –ves, false +ves), and devices' capabilities information.

3. It is capable of auto-generating results in tabular form as well as in charts or graphical form.

5. It also provides the entire test log information as a raw-result-set in form of a text file.

5. It facilitates the test organizers to create and store the batch of test-scenarios prior to the experiment/tests.

6. Currently it has support for Bluetooth and 802.11 networks.

## 4. EVALUATION OF SIMULATOR

We conducted a case study of some of the existing pairing methods using the designed simulator for evaluation purposes. Results of the study have proved to be positive. The objective of the study was to evaluate the usability of four pairing methods as well as the simulator itself. These four methods are Button-to-Button, Display-to-Button, LED-to-Button and Beep-to-Button. First two methods are described in [20], while other two are variations of second method (i.e. Display-to-Button). A brief description of each of the implemented method is given below.

(i) Button-to-Button (B-to-B)

In this method user is required to press and release the button on both of the devices (i.e. device A and device B) simultaneously with random time-intervals. Both of the simulated devices are programmed to start a timer with the first button press. Then, the elapsed time between subsequent button-presses is exploited to calculate the key. From each time-interval 3-bits of the secret are generated.

(ii) Display-to-Button (D-to-B)

In this method target device (device B) selects a key, encode and transmit it through random flashes of the display. Whenever the display of device B flashes user is required to press and release button on device A. Likewise previous approach, the elapsed time between each button-press is used to calculate the bits of shared key on device A.

(iii) LED-to-Button (LED-to-B)

This scheme is similar to Display-to-Button approach. The only difference is that instead of a display, an LED is used to transmit the bits of shared secret. This scheme is suitable in the situations where one of the devices has only a button and the other has only a single LED (e.g. wireless access point). This scheme works in the same way as the previous one. Device B chooses a key and transmits it through LED-blinks with random time-intervals. To obtain same

International Journal of Network Security & Its Applications (IJNSA), Vol.1, No.3, October 2009shared secret on device A, user is responsible to press and release button on the device whenever an LED blinks on device B.

(iv) Beep-to-Button (Beep-to-B)

This scheme is also a variation of Display-to-Button method. In this scheme, device B selects a key and transmits it through random beeps. User is required to press and release button on first device whenever he/she hears a beep sound from the other device. This method is useful in the scenarios where first device has only a single button and the other device has only a speaker.

### 4.1. Test Procedure

A total of 15 volunteers were recruited. All of the participants were chosen on first-come first-serve basis. All of the participants are students and most of them are PhD students. They all are good computer users. The background profile information of the participants is summarized in table 1.

Table 1. Test Participants Profile Information

| Gender | Male: 86.66% |
| --- | --- |
|  | Female: 13.33% |
| Age | 25 – 30: 40% |
|  | 31 – 36: 40% |
|  | 37 – 42: 20% |
| Last academic qualification achieved | Bachelor: 26.66% |
|  | Masters 73.33% |
| Having experience of pairing two devices | Yes: 93.33% |
|  | No: 6.33% |

The tests were conducted in two environments; a lab-based environment using desktop computers, running Windows XP operating system, and a home-based environment using laptops, running Windows Vista operating system. Before the start of each experiment, we have explained briefly the goals of the experiment along with the description of each pairing method to the participant. Then, a pre-test questionnaire is filled by the participant before starting the test cases.

Each experiment consists of three parts. In first part, three methods LED-to-Button, Display-to-Button and Beep-to-Button are tested. Since, the simulator facilitates to generate and store the batch of test scenarios that can be executed later on demand; so, we created a batch of six tests scenarios (two for each method) a priori. This facilitates each participant to perform all of the six tests in one-go without any interruption. In second part of the experiment, each participant performed two repetitions of Button-to-Button method. Left-button of the mouse is used as a button of the simulated device. Participant is required to click on 'Push-Button' simultaneously with random time-intervals on both of the simulated devices. This test requires two machines, one for simulating Alice (button-capable device A) and other for simulating Bob (button-capable device B). In third part, we asked the user to build any of the preferred method using the simulator and execute it. It gave the user an opportunity to examine the usability of the designed simulator. Finally, at the end of experiment every participant filled a post-test questionnaire that contains questions regarding the usability (from very easy to not usable at all) of each of the method and the simulator.

45

International Journal of Network Security & Its Applications (IJNSA), Vol.1, No.3, October 2009

## 4.2. Results and Discussion

Table 2. Results generated by the simulator based on logged data

| Pairing Method | Avg. Completion Time (sec) | Standard Deviation | False Negatives/Safe Errors |
|---|---|---|---|
| LED-To-Button (BEDA Variant-1) | 37.707867 | 17.352257 | 36.666668 |
| BEDA-1 (B-To-B) Button-to-Button | 19.482767 | 4.7960396 | 6.666667 |
| BEDA-2 (D-To-B) Display-to-Button | 33.3115 | 15.323538 | 20.0 |
| Beep-To-Button (BEDA Varient-2) | 35.096397 | 17.274122 | 33.333336 |

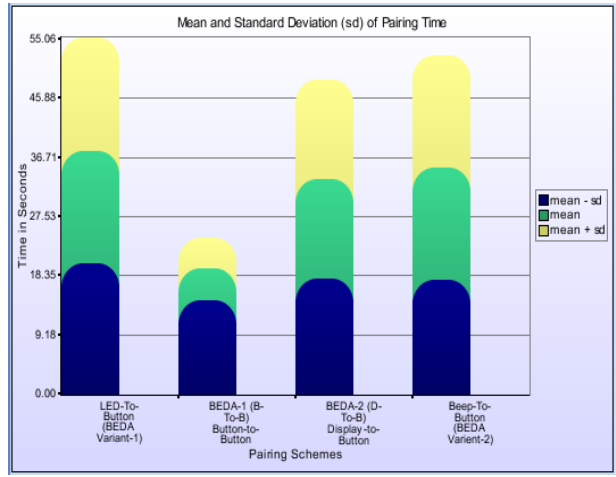

Figure 3: Mean and standard deviation (sd) of pairing time

Results presented in this section are obtained from the collected data by means of questionnaire and as well as by the generated log file of the simulator. Table 2 above shows the results that are auto-generated by the simulator. The graphs shown in figures 3 and 4 are also auto-generated by the simulator using logged data. Other graphs shown in figures 5, 6 and 7 are drawn from the data obtained through post-test questionnaire. Microsoft Excel is used to draw these graphs.

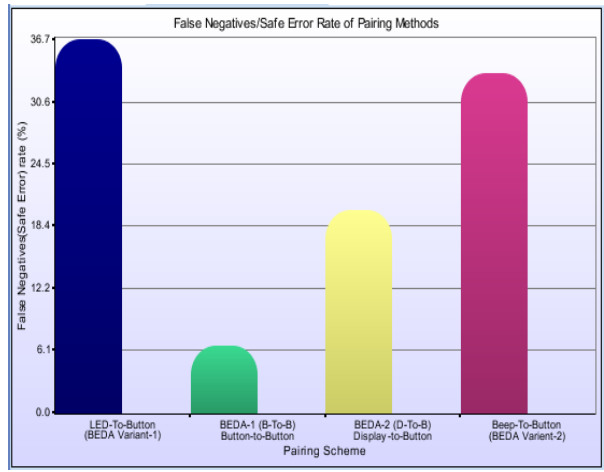

Figure 4: False negatives (safe error) rate of pairing methods

According to the graph shown in figure-3, using a button on both devices is faster than all of the other schemes. Figure 4 shows that B-to-B has the minimum number of false negatives, while LED-to-B has the maximum number of false negatives. Figure 5 shows that majority of the participants considered B-to-B and D-to-B methods very easy to use; while none considered B-to-B method as hard to use. Figure 6 shows that most of the users preferred B-to-B and D-to-B methods over LED-to-B and Beep-to-B methods in the case if their devices support all of the four methods. Graph in figure 7 presents the evaluation results for the designed simulator itself. 33% users considered it very easy to use; while 67% considered it as easy to use and none of them considered it as hard, very hard or not usable at all. These results show that the designed simulator is applicable for testing usability of pairing methods from both developer and users point of view. Since the data presented in table 2 and graphs shown in figures 3 – 7 are self-explanatory, so instead of describing them in more detail, we would like to discuss how this simulator can be helpful in the research of generalizing or standardizing the secure device pairing mechanism.

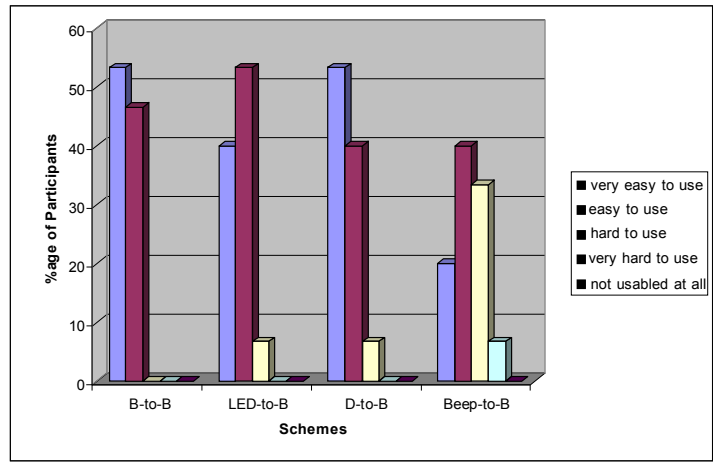

Figure 5: Participants response for the usability of pairing methods

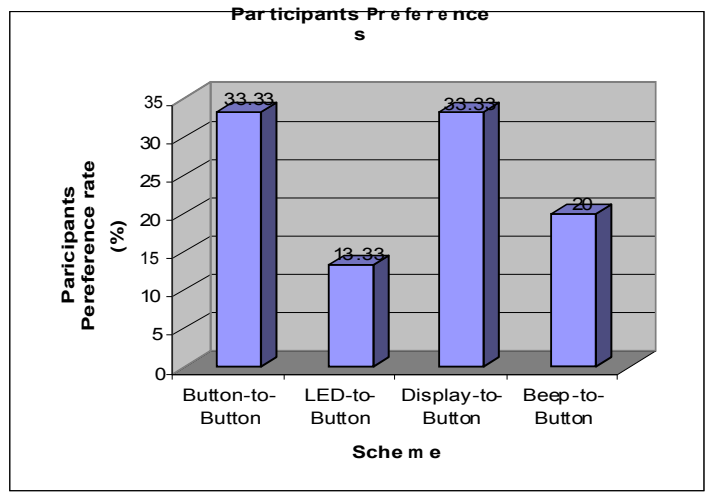

Figure 6: Participants preference of the pairing method

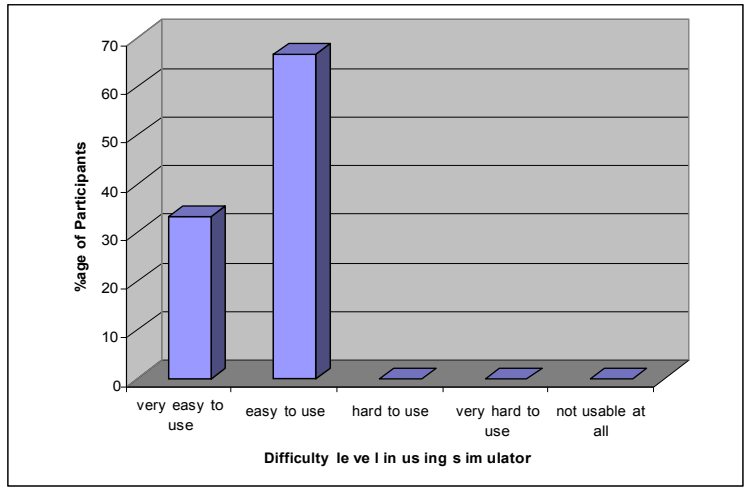

Figure 7: Level of difficulty in using the simulator

As motivation, let's consider a device pairing scenario where device A is only button-capable and device B has display, LED and as well as a speaker. Note that in this scenario we are considering only the four pairing methods along with results obtained as a consequence of our case study. So, in the scenario possible pairing methods are Display-to-Button, LED-to-Button and Beep-to-Button. The questions is how does user (or a system such as [24]) decide the best pairing scheme among three of these possible methods. In this and other similar situations, user or system needs to have the knowledge of priority-level (or some specific measures) of each of the candidate method, so that the best possible pairing method can be selected. This priority-level can be set based on the level of security provided by the method and as well as the usability of that method. We believe that the results of this and other more detailed usability studies using the designed simulator can be helpful to prioritize or set weights for each of the pairing method based on their usability and the level of provided security. For example, based on the obtained results of our case study, Display-to-Button has the highest priority in the scenario presented above. So, user/system should select this protocol to initiate the pairing. Further, this simulator can be helpful to rapidly implement and test a large set of pairing schemes in order to specify certain common or standard measures for up-coming pairing protocols.

## 5. CONCLUSIONS

There has been done an immense work in the field of secure device pairing from both academic research and industrial research points of view. However, no one has yet devised the perfect pairing scheme that could be feasible for all kind of or large set of scenarios. Pairing schemes vary in the strength of their security, the level of required user intervention, their susceptibility to environmental conditions and in the required physical capabilities of the devices as well as the required proximity between the devices. So, we still need other mechanisms, infrastructures, tools and techniques that integrate several pairing techniques within a general architecture for providing secure as well as usable pairing mechanisms (such as [24]). One of the reasons for lack of a general, standardized or universal pairing mechanism is an uneven comparative evaluation of the several existing methods. It might be because of unavailability of specialized tools that provide a common platform to test the usability or security of these methods against some common set of measures. This creates the need to design new tools, such as simulators, benchmarks and usability testing frameworks, that can be used to evaluate the existing as well as new pairing schemes [25]. This motivated us towards the design and development of the pairing simulator. Our designed simulator is capable of generating and saving test scenarios a priori, logging test information and generating textual, tabular as well as graphical results from the logged data at the completion of each test case. We believe that it will be very helpful for both the researchers and other less technical persons working in the area of device pairing to rapidly implement and test new pairing protocols without writing extensive piece of code.

## ACKNOWLEDGEMENTS

This research is sponsored/funded by University of Sindh, Jamshoro, Pakistan under Mega Project Phase-I: No.SU/PLAN/F.SCH/650 and the work presented in this paper is part of authors earlier published work [24, 25].

## REFERENCES

[1] Kumar, A., et al. Caveat eptor: A comparative study of secure device pairing methods. in IEEE International Conference on Pervasive Computing and Communications (PerCom-09). 2009.

[2] Diffie, W. and M.E. Hellman, New Directions in Cryptography. IEEE Transactions on Information Theory, 1976. IT-22(6): p. 644--654.

[3] Stajano, F., The Resurrecting Duckling - What Next?, in Revised Papers from the 8th International Workshop on Security Protocols. 2001, Springer-Verlag.

[4] Stajano, F. and R. Anderson, The Resurrecting Duckling: Security Issues for Ad-hoc Wireless Networks, in Security Protocols. 2000. p. 172-182.

[5] Balfanz, D., et al. Talking to strangers: Authentication in adhoc wireless networks. in Symposium on Network and Distributed Systems Security (NDSS '02). 2002. San Diego, California.

[6] Mayrhofer, R. and M. Welch. A Human-Verifiable Authentication Protocol Using Visible Laser Light. in Availability, Reliability and Security, 2007. ARES 2007. The Second International Conference on. 2007.

[7] Mayrhofer, R., M. Hazas, and H. Gellersen, An authentication protocol using ultrasonic ranging: Technical Report. 2006, Lancaster University.

[8] Mayrhofer, R. and H. Gellersen. On the Security of Ultrasound as Out-of-band Channel. in Parallel and Distributed Processing Symposium, 2007. IPDPS 2007. IEEE International. 2007.

[9] Spahic, A., et al., Pre-Authentication using Infrared Privacy, Security, and Trust Within the Context of Pervasive Computing, 2005.

[10] Shaked, Y. and A. Wool. Cracking the Bluetooth PIN. in MobiSys '05: Proceedings of the 3rd international conference on Mobile systems, applications, and services. 2005. Seattle, Washington: ACM.

[11] Jakobsson, M. and S. Wetzel, Security Weaknesses in Bluetooth, Lecture Notes in Computer Science, 2001. 2020: p. 176+.

[12] McCune, J.M., A. Perrig, and M.K. Reiter, Seeing-is-believing: using camera phones for human-verifiable authentication. Security and Privacy, 2005 IEEE Symposium on, 2005: p. 110 - 124.

[13] Saxena, N., et al., Secure Device Pairing based on a Visual Channel. sp, 2006. 0: p. 306-313.

[14] Goodrich, M.T., et al. Loud and Clear: Human-Verifiable Authentication Based on Audio. in Distributed Computing Systems, ICDCS 2006. 26th IEEE International Conference. 2006.

[15] Soriente, C., G. Tsudik, and E. Uzun (2007) HAPADEP: Human Asisted Pure Audio Device Pairing. Cryptology ePrint Archive, Report 2007/093.

[16] Holmquist, L.E., et al., Smart-Its Friends: A Technique for Users to Easily Establish Connections between Smart Artefacts, in Proceedings of the 3rd international conference on Ubiquitous Computing. 2001, Springer-Verlag: Atlanta, Georgia, USA.

[17] Mayrhofer, R. and H. Gellersen, Shake Well Before Use: Authentication Based on Accelerometer Data, in 5th International Conference on Pervasive Computing (Pervasive 2007). 2007.

[18] Castelluccia, C. and P. Mutaf, Shake them up!: a movement-based pairing protocol for CPU-constrained devices, in Proceedings of the 3rd international conference on Mobile systems, applications, and services. 2005, ACM: Seattle, Washington.

[19] Varshavsky, A., et al., Amigo: Proximity-Based Authentication of Mobile Devices, in UbiComp 2007: Ubiquitous Computing. 2007. p. 253-270.

[20] Soriente, C., G. Tsudik, and E. Uzun. BEDA: Button-Enabled Device Association, in Internation Workshop on Security and Spontaneous Interaction (IWSSI 2007). 2007.

[21] Nicholson, A., et al., LoKey: Leveraging the SMS Network in Decentralized, End-to-End Trust Establishment, in Pervasive Computing. 2006. p. 202-219.

[22] Lester, J., B. Hannaford, and G. Borriello. Are You with Me?" - Using Accelerometers to Determine If Two Devices Are Carried by the Same Person. in Pervasive Computing. 2004: Springer-Verlag (2004) pg. 33-50.

[23] Malkani, Y.A., D. Chalmers, and I. Wakeman. Towards a General System for Secure Device Pairing by Demonstration of Physical Proximity (Poster). in UBICOMP Grand Challenge: Workshop on Ubiquitous Computing at a Crossroads: Art, Science, Politics and Design. 6th and 7th January, 2009. Huxley Building, Imperial College, London.

[24] Malkani, Y.A., et al., Towards a General System for Secure Device Pairing by Demonstration of Physical Proximity, in MWNS-09 co-located with IFIP Networking 2009 Conference, Shaker Verlag: Aachen, Germany. ISBN: 978-3-8322-8177-9. p. 13-24.

[25] Malkani, Y.A. and L.D. Dhomeja. Secure Device Association for Ad Hoc and Ubiquitous Computing Environments. in 5th IEEE International Conference on Emerging Technologies (ICET-09), 2009.

[26] Saxena, N. and J. Voris. Pairing Devices with Good Quality Output Interfaces. in International Workshop on Wireless Security and Privacy (WISP) (co-located with ICDCS). . 2009.

[27] The Legion of the Bouncy Castle, Url:http://bouncycastle.org/java.html.

[28] The Chart2D Tutorial, URL:http://chart2d.sourceforge.net/Chart2D_1.9.6/Tutorial/Tutorial.htm.

**Authors**

**Mr. Yasir Arfat Malkani** is Lecturer at the Institute of Mathematics and Computer Science (IMCS), University of Sindh, Jamshoro, Pakistan. Currently, he is a DPhil student and Associate Tutor at University of Sussex, UK. He was awarded Vice Chancellor's silver medal for obtaining first position in M.Sc Computer Science at the University of Sindh, Jamshoro (Pakistan) in year 2003. He was appointed as a Research Associate in University of Sindh in 2004, and then as a Lecturer in July 2005. He was awarded PhD Scholarship from University of Sindh in year 2006 to pursue his DPhil studies at University of 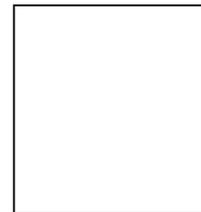 Sussex, Brighton, UK. His main area of research is Pervasive Computing, and his research is focused on context awareness and security issues in pervasive and ubiquitous computing environments. Currently, he is working on the design and development of a generic framework for secure pairing of pervasive devices by demonstration of physical proximity.

**Mr. Lachhman Das Dhomeja** is Assistant Professor at the Institute of Information Technology (IIT), University of Sindh, Jamshoro, Pakistan. Currently, he is a DPhil student at University of Sussex, Brighton, UK. He got his Master's degree in Computer Technology from University of Sindh, Jamshoro (Pakistan). He was awarded PhD Scholarship from University of Sindh in year 2006 to pursue his DPhil studies at University of Sussex, Brighton, UK. His main research interest is in the area of Pervasive Computing. Currently, he is working on policy-based 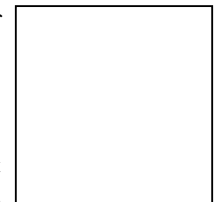 adaptive systems for Pervasive Computing environments.